\documentclass{aa}
\usepackage{latexsym}
\usepackage{epsfig}
\newcommand{\radm}{rad~m$^{-2}$} 
\newcommand{\pl}{\parallel}

\newcommand{\srm}{\mbox{$\sigma_{\scriptscriptstyle\rm RM}$}}
\newcommand{\dg}{\mbox{$^{\circ}$}}
\begin{document}

\title{Structure in the polarized Galactic synchrotron 
       emission, in particular ``depolarization canals''}

   \author{M. Haverkorn\inst{1}
          \and
          P. Katgert\inst{2}
	  \and
	  A. G. de Bruyn\inst{3,4}
          }
   \offprints{M. Haverkorn}
   \institute{Leiden Observatory, P.O.Box 9513, 2300 RA Leiden, the
              Netherlands\\
	      (Current address: Harvard-Smithsonian Center for
	      Astrophysics, 60 Garden Street MS-67, Cambridge MA
	      02138, USA)\\
              \email{mhaverkorn@cfa.harvard.edu} 
         \and Leiden Observatory, P.O.Box 9513, 2300 RA Leiden, the
              Netherlands\\ 
              \email{katgert@strw.leidenuniv.nl} 
         \and ASTRON, P.O.Box 2, 7990 AA Dwingeloo, the Netherlands\\ 
              \email{ger@astron.nl}
         \and Kapteyn Institute, P.O.Box 800, 9700 AV Groningen,
              the Netherlands }
\date{Received, accepted}

\abstract{ 
  The polarized component of the diffuse radio synchrotron emission
  of our Galaxy shows structure, which is apparently unrelated to the
  structure in total intensity, on many scales. The structure in the
  polarized emission can be due to several processes or
  mechanisms. Some of those are related to the observational setup,
  such as beam depolarization -- the vector combination and (partial)
  cancellation of polarization vectors within a synthesized beam --,
  or the insensitivity of a synthesis telescope to structure on large
  scales, also known as the 'missing short spacings problem'. Other
  causes for structure in the polarization maps are intrinsic to the
  radiative transfer of the emission in the warm ISM, which induces
  Faraday rotation and depolarization.

  We use data obtained with the Westerbork Synthesis Radio Telescope
  at 5~frequencies near 350~MHz to estimate the importance of the
  various mechanisms in producing structure in the linearly polarized
  emission. In the two regions studied here, which are both at
  positive latitudes in the second Galactic quadrant, the effect of
  'missing short spacings' is not important. The properties of the
  narrow depolarization 'canals' that are observed in abundance lead
  us to conclude that they are mostly due to beam depolarization, and
  that they separate regions with different rotation measures. As beam
  depolarization only creates structure on the scale of the
  synthesized beam, most of the structure on larger scales must be due
  to depth depolarization. We do not discuss that aspect of the
  observations here, but in a companion paper we derive information
  about the properties of the ISM from the structure of the polarized
  emission.
    }

\titlerunning{Structure in the polarized synchrotron background}
\maketitle

\section{Introduction}
\label{s3:intro}

The omnipresent cosmic rays in the Milky Way, spiraling in the
Galactic magnetic field, provide a synchrotron radio background which
is partially polarized. This radiation propagates through the warm
ionized interstellar medium (ISM) and is modulated by it. This makes
observations of the polarized continuum radio background a valuable
tool for studies of the warm ISM and the Galactic magnetic field.

Generally, observations of the polarized synchrotron emission of the
Galaxy show small-scale structure in polarized intensity $P$\/ or
polarization angle $\phi$ uncorrelated with structure in total
intensity $I$ (e.g.\ Wieringa et al.\ 1993, Duncan et al.\ 1999, Gray
et al.\ 1999, Gaensler et al.\ 2001, Uyan\i ker and Landecker 2002 for
the Milky Way; Horellou et al.\ 1992,  Berkhuijsen et al.\ 1997 for
M51; Shukurov and Berkhuijsen 2003, Fletcher et al.\ 2004 for
M31). The lack of  correlation between $P$ and $I$ indicates that the
structure in polarization is not exclusively due to intrinsic
structure in  synchrotron emission.  Instead, the fluctuations in
polarization angle are explained in terms of Faraday rotation of the 
synchrotron radiation that impinges on the magneto-ionic medium of the
ISM relatively close to the Sun (Burn 1966). Multi-frequency
polarimetry of the synchrotron emission allows determination of the
Faraday rotation  measure $RM \propto \int n_e B_{\pl} \, ds$, which
depend on electron density $n_e$, magnetic field parallel to the line
of sight $B_{\pl}$ and path length $ds$. Thus, study of RMs enables
the study of the structure and electron-density-weighted strength of
the Galactic magnetic field.

However, whereas Faraday rotation can explain the
variation in polarization angle, it does not provide an explanation
for the structure in polarized intensity $P$. Although the lack of
zero-baseline visibilities in some interferometric observations could
produce structure in $P$\/ from structure in $\phi$, this would leave
the structure in $P$\/ in absolutely calibrated single-dish
observations unaccounted for. Therefore, depolarization must also
contribute to structure in polarized intensity and polarization
angle. Detailed analysis of the different depolarization mechanisms
yield unique information on the magnetic field.

Two different approaches to the description of depolarization
mechanisms can be found in the literature: one which is based on the
physical processes that produce the depolarization, and another that
makes a geometrical distinction between effects in depth and in angle.
In this paper, we use the latter, which is more convenient for our
purpose. We will first discuss the physical processes causing
depolarization, and then describe how these are treated here. For
extensive treatments of the depolarization processes see e.g. Gardner
and Whiteoak (1966), Burn (1966), or Sokoloff et al.\ (1998).

\begin{itemize} 
\item {\em Wavelength independent depolarization} is due to turbulent
      magnetic fields in the ISM. Cosmic rays in a turbulent magnetic
      field emit synchrotron radiation with varying polarization
      angle. Therefore, superposition of the polarization vectors
      along the line of sight and across the telescope beam results in
      partial depolarization of the emission, independent of
      wavelength. No Faraday rotation is involved.
\item {\em Differential Faraday rotation} occurs if a medium
      contains thermal and relativistic electrons and a (partly)
      regular magnetic field. Synchrotron radiation emitted at
      different distances along the line of sight undergo  different
      amounts of Faraday rotation. This is a one-dimensional (along
      the line of sight), wavelength dependent depolarization effect.
\item {\em Internal Faraday dispersion} is the depolarization in a
      turbulent synchrotron-emitting magneto-ionic medium. It is
      a combination of the two effects described above, but it also
      involves depolarization within the telescope beam. Variation
      in intrinsic polarization angle and in Faraday rotation, which
      occur both along the line of sight and across the telescope
      beam, cause depolarization of the radiation.
\end{itemize}

Here, it is more convenient to divide these depolarization mechanisms
in those operating along the line of sight and those perpendicular to
the line of sight, regardless of the physical process causing the
depolarization. The latter mechanism is called {\em beam
depolarization} (Gray et al.\ 1999, Gaensler et al.\ 2001, Landecker
et al.\ 2001, which is a result of vector averaging for neighboring
directions within the same telescope beam. Depolarization along the
line of sight is {\em depth depolarization} (Landecker et al.\ 2001,
Uyan\i ker and Landecker 2002, referred to as front-back
depolarization by Gray et al.\ 1999), and is a one-dimensional
addition of polarization vectors, assuming an infinitely narrow
telescope beam. Depth depolarization is a combination of
differential Faraday rotation, internal Faraday dispersion and
depolarization due to variations of the intrinsic polarization angle
along  the line of sight.

In this paper, we discuss various processes that can produce structure
in $P$, and we use several multi-frequency datasets obtained with the
Westerbork Synthesis Radio Telescope (WSRT) to gauge their
importance. The influence of missing short spacings and of
depolarization mechanisms on the data is estimated, as well as the
importance of beam depolarization. A discussion of the effects of
depth depolarization is given in a companion paper (Haverkorn et
al. 2004). 

In Sect.~\ref{s3:obs} we summarize the relevant parameters of the
Westerbork polarization observations that will be used to estimate the
importance of the various effects that contribute to the structure in
$P$. In Sect.~\ref{s3:lss} the r\^ole of missing short spacings in
interferometer measurements is estimated, and we discuss how the
resulting images can be affected. Section~\ref{s3:can} presents a
discussion on the origin of depolarization canals in polarized
intensity. Finally, the conclusions are stated in
Sect.~\ref{s3:conc}. 

\section{The observations}
\label{s3:obs}

\begin{table}
  \begin{center}
    \begin{tabular}{l|cc}
                & Auriga           & Horologium \\
      \hline
      (l,b)     & \hspace{.5cm}(161\dg, 16\dg)\hspace{.5cm}  & (137\dg, 7\dg) \\
      size      & 7\dg$\times$9\dg & 7\dg$\times$7\dg \\
      FWHM      &5.0\arcmin$\times$6.3\arcmin&5.0\arcmin$\times$5.5\arcmin\\
      pointings & 5$\times$7       & 5$\times$5 \\
      Noise (mJy/bm) & $\sim$~4 (0.5~K) & $\sim$~5 (0.7 K)\\
      1 mJy/bm = & 0.127 K  & 0.146 K \\
      $\;\;\;\;$(at 350 MHz) & & \\
      bandwidth & 5 MHz            & 5 MHz \\
      frequencies & \multicolumn{2}{c}{341, 349, 355, 360, 375 MHz}\\
      \hline
    \end{tabular}
    \caption{Details of the WSRT polarization observations in the
             constellations Auriga and Horologium.}
    \label{t3:data}
  \end{center}
\end{table}

We use Westerbork Synthesis Radio Telescope (WSRT) observations
around 350~MHz in two fields in the constellations Auriga and
Horologium, which were described in detail by Haverkorn
et al.\ (2003a, 2003b). Some details of the observations are
described in Table~\ref{t3:data}.

\begin{figure*}
  \begin{center}
    \hbox{\psfig{figure=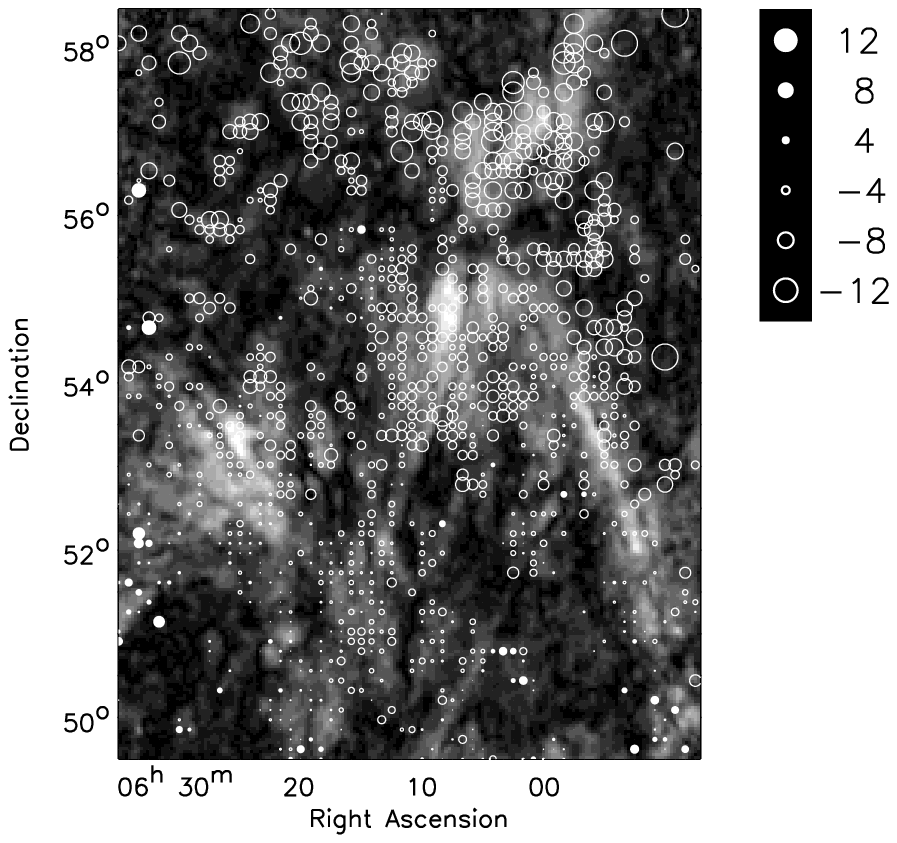,width=.47\textwidth}
          \psfig{figure=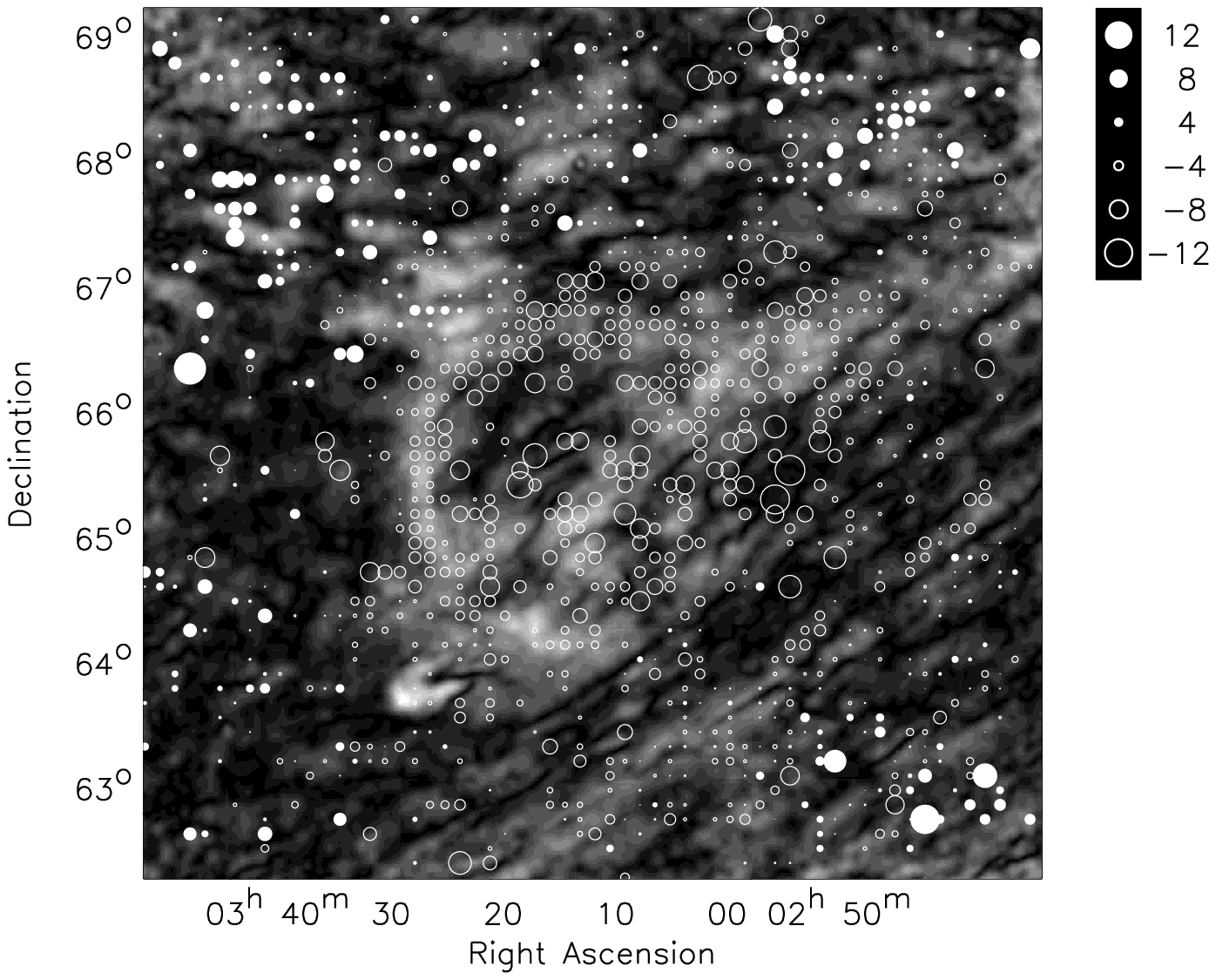,width=.52\textwidth}}
    \caption{Rotation measure maps of the regions in Auriga
      (left) and in Horologium (right), superimposed on polarized
      intensity in grey scale where the maximum is 90 mJy/beam
      for Auriga and 110 mJy/beam for Horologium. Rotation measures
      are denoted by white circles, where filled circles are positive
      $RM$s. The diameter of the symbol represents the magnitude of
      $RM$, and the scaling is given in \radm. Only $RM$s for which
      $\langle P \rangle \ge 5\sigma$ and reduced $\chi^2$ of the
      linear $\phi(\lambda^2)$-relation $< 2$ are shown, and only
      every second independent beam.}
    \label{f3:rmmap}
  \end{center}
\end{figure*}

Fig.~\ref{f3:rmmap} shows the RM distribution in the Auriga (left) and
Horologium (right) fields as circles, superposed on polarized
intensity in grey scale. The structure in $P$ is uncorrelated with
total intensity $I$. The RMs were derived from $\phi = \phi_0 +
RM\lambda^2$, where $\phi_0$ is the intrinsic polarization angle at
$\lambda = 0$. Depolarization mechanisms alter Stokes $Q$ and $U$, and
can destroy the linear $\phi(\lambda^2)$-relation, in which case the
determined $RM$ will not have its simple meaning, viz. $\int n_e
B_{\parallel} ds$ (Sokoloff et al.\ 1998). Therefore, we only show
RM values at positions where (a) the reduced $\chi^2$ of the linear
$\phi(\lambda^2)$-relation $\chi^2_{red} < 2$, and (b) the polarized
intensity averaged over frequency $\left<P\right> > 20$~mJy/beam
($\sim4$ to 5~$\sigma$). The upper limit of $\chi^2 = 2$ is
chosen to allow for slight non-linearity due to depolarization.

\section{The effect of missing short spacings in aperture synthesis 
         observations}
\label{s3:lss}

In aperture synthesis observations, structure on large angular scales
is not well represented because visibilities cannot be measured on
baselines smaller than the diameter of the primary elements. (Also
single-dish observations miss information about structure on scales
larger than the mapped region, and missing flux must be added from
absolutely calibrated polarization maps.) The shortest baseline of the
WSRT is 36m, so that at 350~MHz, structure on angular scales larger
than about a degree is not adequately measured. The proper way to
correct for this undetectable large-scale structure is to observe the
same region at the same frequencies with a single-dish telescope with
absolute intensity scaling and add these large-scale data to the
interferometer data (see Stanimirovic (2002) for methods of data
addition, and Uyan\i ker et al.\ (1998) who have first done this for
diffuse polarization data). However, for the WSRT at 350 MHz this is
not possible, as there is no instrument of suitable size operating at
these frequencies. 

In the data reduction process of the WSRT, the lack of information on
scales larger than about a degree is dealt with by setting the average
value of measured intensities on the scale of the whole field to
zero. For a strong source this will result in an image which has a
bowl-like depression around the source, but for approximately uniform
diffuse emission, it produces a more or less constant offset.  In the
case of polarimetry, this means that the average Stokes $Q$ and $U$
components are set to zero.  So in each observed $Q$ and $U$ map there
may be constant offsets $Q_0$ and $U_0$ that have to be added to the
observed $Q$ and $U$ to obtain the real linearly polarized signal on
the sky. Since the large mosaics are produced from several tens of
pointings, each of which can have its own offsets, the offsets can
vary over a mosaic.

\subsection{The effect of offsets on polarized intensity and rotation 
            measure}

\begin{figure*}
  \begin{center}
    \psfig{figure=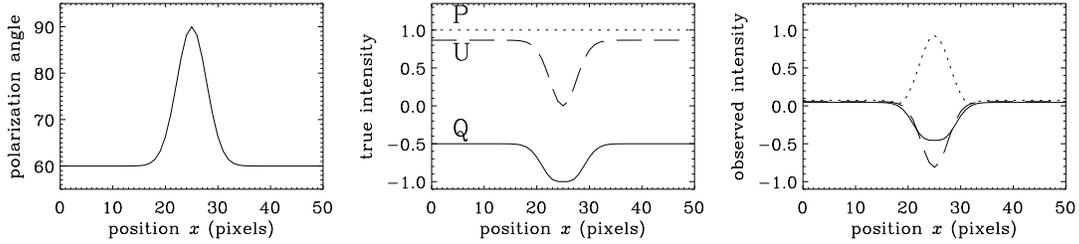,width=.8\textwidth}
    \caption{Illustration of how offsets can cause small-scale
      structure in $P$. Left: an model polarization angle distribution
      in one dimension, where $P$ is assumed constant. Center:
      small-scale structure in $Q$ (solid line) and $U$ (dashed line)
      corresponding to the change in polarization angle, while $P$
      (dotted line) remains constant. Right: the interferometer
      response to this distribution, where $P$ does show apparent
      structure.}
    \label{f3:qu0struc}
  \end{center}
\end{figure*}

The presence of offsets can create spurious small-scale structure in
observed $P$. In particular, offsets can create additional
depolarization canals (see Sect.~\ref{s3:can}) as shown in Uyan\i ker
et al.\ (1998). Fig.~\ref{f3:qu0struc} shows an example of the effect
of offsets on $Q$ and $U$. The left plot gives a simple
one-dimensional model of a change in polarization angle, which causes
small-scale structure in the distribution of $Q$ and $U$, but not in
$P$ (center plot).  The right plot shows the response of an
interferometer: the average $Q$ and $U$ over the field are subtracted
from the signal on the sky. $P_{obs}$ which is computed from $Q_{obs}$
and $U_{obs}$ does show apparent structure on small scales, although
in reality it does not have that structure.

\begin{figure*}
  \begin{center}
    \psfig{figure=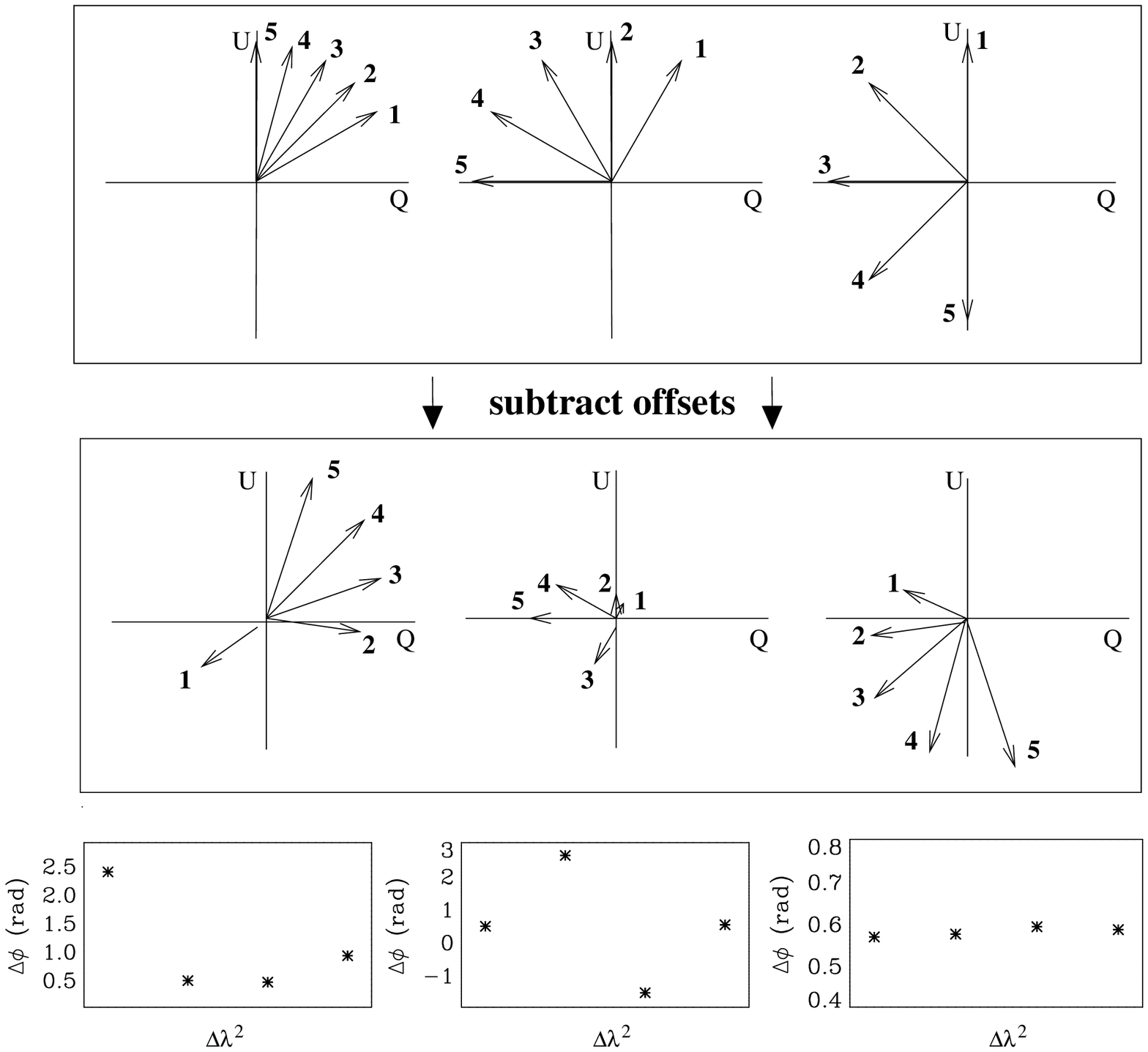,width=.6\textwidth}
    \caption{The effect that offsets have on apparent RMs. Top:
      hypothetical polarization vectors at 3 adjacent positions in the
      sky for three values of true RM, at 5 wavelengths denoted
      1 to 5. Center: results after subtracting offsets determined
      from the situation in the top panel, showing how the offsets can
      destroy the linear $\phi(\lambda^2)$-relation. Bottom: 4
      estimates of $\Delta\phi$ for each wavelength step
      $\Delta\lambda^2$, which gives the apparent $RM =
      \Delta\phi/\Delta\lambda^2$ as the slope of a linear fit of
      $\phi$ to $\lambda^2$.}
    \label{f3:qu0rm}
  \end{center}
\end{figure*}

Furthermore, if undetected offsets are present in the data, the
polarization angle computed from the detected $Q$ and $U$ will, in
general, not show a linear dependence on $\lambda^2$. Therefore, the
fitted RM will then differ from the real one. Although in
interferometer observations the Stokes $Q$ and $U$ emission can be
separated in (observable) small-scale structure and (unobservable)
large-scale structure, this is in general {\it not} true for the
rotation measure, due to the complicated non-linear relation between
intensity at different frequencies and $RM$. Therefore, large-scale
structure or constant non-zero RMs can be observed as long as
small-scale structure in RM is present as well to cause sufficient
structure in $Q$ and $U$ on small scales.

A simple example of how offsets may destroy the linear
$\phi(\lambda^2)$-relation and result in erroneous determinations of
$RM$ is given in Fig.~\ref{f3:qu0rm}. Six plots in the $(Q, U)$-plane
are shown, each with five polarization vectors ${\mathbf P} =
P\exp(-2i\phi)$.  Each vector refers to one of the five wavelengths,
which are equally spaced in $\lambda^2$, and are numbered according to
increasing $\lambda^2$. The upper three plots give a hypothetical
situation of three values of true RM at three adjacent positions,
where the vectors denote the true polarization. All three $RM$s are
chosen to be positive, and the value of $RM$ in the left plot is
doubled and tripled in the central and right hand plot,
respectively. Of these three plots, the $Q$ and $U$ values averaged
over the three positions for each band separately were subtracted
after which the polarization vectors in the lower plots were
obtained. Below these are shown the resulting values of $\Delta\phi$
between two adjacent wavelengths, which gives the apparent $RM =
\Delta\phi / \Delta\lambda^2$. The linear $\phi(\lambda^2)$-relation
is thoroughly destroyed, and the apparent RM deviates from the true
$RM$. 

The fact that offsets can create structure in $P$ and prohibit
reliable $RM$ determinations has to be given serious consideration in
all interferometer observations with missing short spacings.  We will
estimate the importance of offsets in our observations in the next
Section.

\subsection{The importance of offsets in the observations}
\label{ss3:depth}

\begin{table}
  \begin{center}
    \begin{tabular}{r|ccccc|c}
      \multicolumn{1}{c}{($\delta$)} & \multicolumn{5}{c}{\bf Auriga} & \\
        \cline{2-6}
      56.2\dg & 3.76 & 3.11 & 2.78 & 2.28 & 3.14 & \\
      55.0\dg & 3.73 & 3.64 & 3.11 & 2.68 & 3.15 & \\
      53.7\dg & 2.94 & 2.47 & 1.51 & 1.82 & 3.34 & \\
      52.5\dg & 1.99 & 1.88 & 1.90 & 2.00 & 2.82 & \\
      51.2\dg & 2.60 & 1.87 & 1.93 & 2.21 & 2.37 & \\
      50.0\dg & 2.26 & 1.99 & 1.79 & 2.46 & 2.80 & \\ 
      48.7\dg & 3.09 & 2.74 & 2.57 & 2.52 & 3.32 & \\
            \cline{2-6}
      \multicolumn{1}{c}{\mbox{}} & 96.6\dg & 94.6\dg & 92.5\dg 
       & 90.5\dg &\multicolumn{1}{c}{88.4\dg} & ($\alpha$)\\
      \multicolumn{7}{c}{\mbox{}}\\
      \multicolumn{1}{c}{($\delta$)} & \multicolumn{5}{c}{\bf Horologium} \\
      \cline{2-6}
      68.5\dg & 2.19 & 2.59 & 1.58 & 3.12 & 2.71 & \\
      67.3\dg & 2.68 & 2.01 & 1.83 & 1.57 & 1.66 & \\
      66.0\dg & 4.42 & 1.74 & 1.96 & 2.04 & 2.41 & \\
      64.8\dg & 2.72 & 1.84 & 1.80 & 1.94 & 2.16 & \\
      63.5\dg & 2.35 & 1.78 & 1.85 & 1.32 & 1.91 & \\
      \cline{2-6}
        \multicolumn{1}{c}{\mbox{}} & 54.4\dg & 51.1\dg & 48.0\dg
      & 44.9\dg & \multicolumn{1}{c}{41.9\dg}  & ($\alpha$)\\
    \end{tabular}
    \caption{Width of $RM$ distribution for each pointing in Auriga
     (top) and Horologium (bottom). The numbers denote
    \srm\ in \radm, for the pointings given.}
    \label{t3:srm}
  \end{center}
\end{table}

\begin{figure*}
  \centering
  \psfig{figure=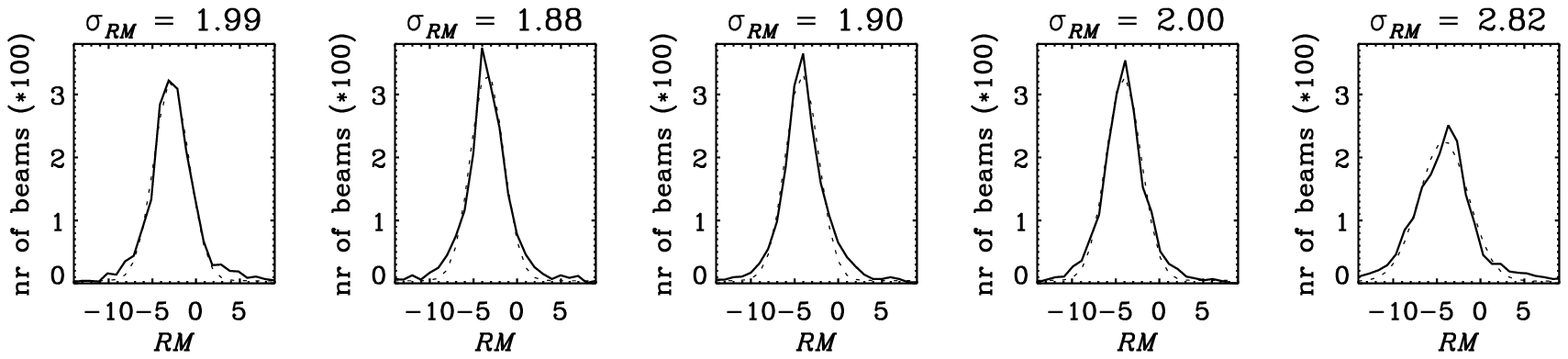,width=.9\textwidth}
  \caption{$RM$ distributions in separate pointings. The plots show
           the central row of pointings in the Auriga region. Dotted
           lines are Gaussian fits to the data, and the fitted
           $\sigma_{RM}$ are given above the plot.}
  \label{f3:poi}
\end{figure*}

The offset in each of the pointings of a mosaic depends on the spread
in RM in that pointing, see Appendix~\ref{a3:screen}. Therefore, we
determined \srm\ for each pointing in the two mosaics. The values of
\srm\ for each pointing position in Auriga and Horologium are given in
Table~\ref{t3:srm}, while Fig.~\ref{f3:poi} shows an example of the
$RM$ distributions in the central row of pointings in the Auriga field.

\begin{figure*}
  \centerline{\psfig{figure=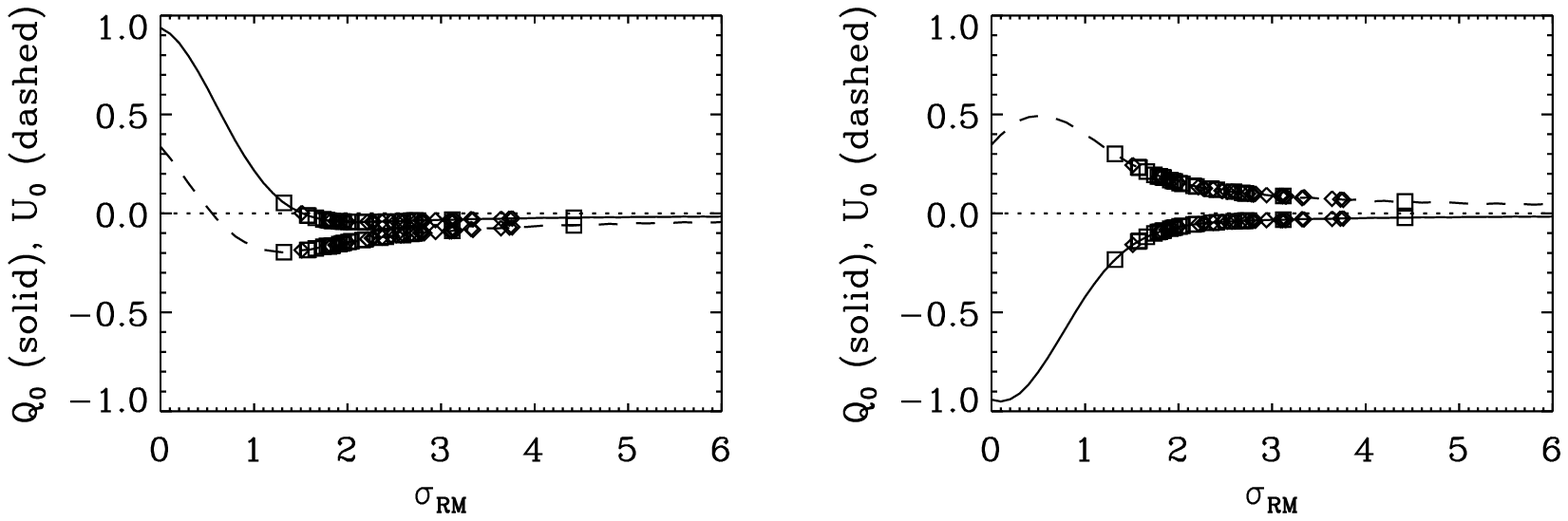,width=.8\textwidth}}
  \caption{Missing large-scale structure in $Q$ (solid line) and $U$
  (dashed line) for a Faraday screen with a Gaussian RM distribution
  with width $\sigma_{RM}$ and $P_0= 1$, assuming that the offsets are
  constant over the field. Diamonds denote $\sigma_{RM}$ observed in
  the pointings of the Auriga field, squares are the pointings of the
  Horologium field. Offsets depend on the intrinsic polarization angle
  $\phi_0$, shown are the best- and worst-case scenario for $\phi_o =
  10\dg$ (left) and $\phi_0 = 80\dg$ (right).}
  \label{f3:qu0}
\end{figure*}

For polarized radiation traveling through a non-emitting Faraday
screen with a Gaussian random distribution of $RM$s, offsets can be
described as (see Appendix~\ref{a3:screen})
\begin{eqnarray}
  Q_0  &=& A \left[ \cos(2\phi_0) - \frac{1}{\sqrt{\pi}}
             \sin(2\phi_0)\;F\right] \label{e:q0}\\
  U_0  &=& A \left[ \sin(2\phi_0) - \frac{1}{\sqrt{\pi}}
             \cos(2\phi_0)\;F \right] \label{e:u0}\\
  \mbox{where}\;F&=& \int_0^{\sqrt{2}\srm\lambda^2} \mbox{e}^{t^2} dt 
             \;\;\;\mbox{and}\;\;\;
	     A= P_0 \mbox{e}^{-2\srm^2\lambda^4} \nonumber
\end{eqnarray}
and offsets are assumed constant over the pointing, with $\phi_0$ the
intrinsic polarization angle and $P_0$ the initial polarized
intensity. This confirms that high RM dispersion randomizes any
uniform polarized background so that no large-scale components in $Q$
and $U$ remain. Fig.~\ref{f3:qu0} displays the offsets as calculated
from Eqs.~(\ref{e:q0}) and~(\ref{e:u0}) for each pointing. The solid
and dashed lines denote the theoretical $Q_0$ and $U_0$ respectively,
the symbols show the \srm\ in the Auriga field (diamonds) and in the
Horologium field (squares). As offsets also depend on $\phi_0$, the
figure shows minimal offsets (for $\phi_0 = 10\dg$, left) and maximal
offsets (for $\phi_0 = 80\dg$, right). The symbols denote the
$\sigma_{RM}$ for pointings in the Auriga field (diamonds) and
Horologium field (squares).

In the best case ($\phi_0 = 10\dg$), no pointings in Auriga and one
pointing in Horologium have offsets exceeding 20\%, which corresponds
to an error in polarization angle $\varepsilon_{\phi}$ of 10\%. In the
worst case ($\phi_0 = 80\dg$), 3 pointings in Auriga (out of 35) and 9
in Horologium (out of 25) allow offsets above 20\%. However, only one
pointing (in Horologium) would allow offsets higher than 27\%
($\varepsilon_{\phi}=14$\%). Therefore, if our observations result
from a uniformly polarized background viewed through a
Faraday-modulating screen, the $\sigma_{RM}$ in most pointings in our
fields is so large that missing large-scale structure leads to an
error in polarization angle of less than 10\%.

\begin{figure}[t]
  \psfig{figure=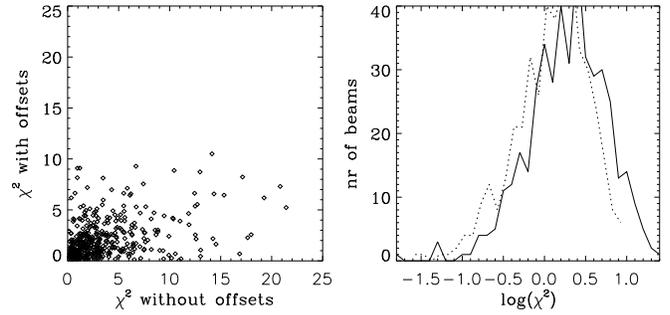,width=.48\textwidth}
  \caption{The influence of offsets on reduced $\chi^2$ in a subfield
    of Auriga. Left: reduced $\chi^2$ distribution of linear
    $\phi(\lambda^2)$-relation with best-fit offsets against values of
    reduced $\chi^2$ without offsets. Right: distribution of reduced
    $\chi^2$ with computed offsets (dotted line) and without offsets
    (solid line).}
  \label{f3:chi}
\end{figure}

The magnitude of the offsets can also be estimated from the variation of
observed polarization angle $\phi$ with wavelength. The observations
show that $\phi$ does not perfectly follow the linear relation $\phi =
\phi_0 + RM \lambda^2$, as one would expect for pure Faraday rotation. 
Offsets in $Q$ and/or $U$ can cause deviations in the linear
$\phi(\lambda^2)$-relation, so we estimated in both fields the offsets
that would minimize the observed non-linearities in the
$\phi(\lambda^2)$-relation.

For this, subfields of $\sim1\dg\times1\dg$ were selected around a
pointing center. A large-scale constant $Q_0$ and/or $U_0$, 
independent for each frequency, were added to the data to minimize
the $\chi^2$ of the $\phi(\lambda^2)$-relation. Resulting offset values
in some subfields with a magnitude of the same order as $P$ in that
field, which decreased the average $\chi_{red}^2$ by a factor of
two. However, Fig.~\ref{f3:chi} shows the distribution of individual
reduced $\chi^2$ values per beam for a typical pointing in the Auriga
field. The left plot shows reduced $\chi^2$ values computed with
offsets against reduced $\chi^2$ without offsets. In the right hand
panel, the histograms of $\chi^2$ without (solid line) and with offsets
(dotted line) are given. Offsets only cause a decrease in
$\chi^2$ in 52\% of the beams, although the average $\chi^2$
diminishes. In other pointings, this percentage ranges from 
49\% to 71\%.  Therefore, the computed offsets do not give a real
improvement of the data, and cannot be considered real missing
large-scale components. Of course, this argument assumes that offsets
are the only agents distorting the linear $\phi(\lambda^2)$-relation,
while depolarization mechanisms can yield non-linearity too. In
addition, it assumes that the offsets are constant over the
subfield considered, which may not be true either. Probing smaller
subfields is no solution for this problem as the number of data points
becomes too small with respect to the number of free parameters.

A third argument against dominant offsets in the data is the high
quality of the determination of $RM$, i.e. a linear
$\phi(\lambda^2)$-relation with a low $\chi^2$. 
Of all pixels with high enough polarized intensity ($\left<P\right> >
20$~mJy/beam), $\sim$~70\% (in Auriga) and $\sim$~62\% (in Horologium)
has a reduced $\chi^2 < 2$. If offsets of the same order of the data would
exist, $RM$s could not be so well-determined over such a large part of
the fields. For ideal data with constant $P$, random offsets
  cause a $\chi^2 > 2$ if the offsets are larger than $\sim$~8\%.

Finally, models of depolarization in a synchrotron-emitting and
Faraday-rotating medium, which are presented in a companion paper
(Haverkorn et al. 2004a), do not show average $Q$ or $U$ values $\ga
10$~mJy/beam (2$\sigma$).
 
From the large \srm, the good quality of the $RM$ determinations,
the depolarization models, and from solving for offsets that minimize
$\chi^2$, we conclude that the presence of considerable undetected
large-scale structure due to the missing short spacings is unlikely. 

This conclusion can also be checked with absolutely calibrated polarized
intensity maps at 408~MHz by Berkhuijsen and Brouw (1963). This
frequency is close enough to 350~MHz to allow comparison, although the
polarized intensity at 408~MHz is expected to be slightly higher
because the polarization horizon is further away at this frequency. We have
smoothed our data to the 2\dg\ FWHM of Berkhuijsen and Brouw, and
derived any missing large-scale structure by comparing the two data
sets. The polarized brightness temperatures at 408~MHz at the
positions of the Auriga and Horologium fields are 1.8~K and 2.7~K, 
respectively. Using a power law spectral index of 2.7, this
corresponds to 2~K and 3~K at 350~MHz. The polarized brightness
temperatures derived from the smoothed data are 0.07~K and 0.12~K,
respectively. Converting from Kelvin to Jansky~per~beam (see
Table~\ref{t3:data}) and taking into account that offsets in $Q$ and
$U$ are on average a factor $\sqrt{2}$ smaller than those in $P$, this
means that any missing large-scale components in Stokes $Q$ and $U$
are smaller than 10.6~mJy~beam$^{-1}$ for the Auriga region, and
13.7~mJy~beam$^{-1}$ for Horologium. For both fields, this corresponds
to about 2 to 3 signal-to-noise in $Q$ and $U$, although it is not
known what the influence of the difference in polarization horizon
is. We conclude that these data are not in disagreement with our
conclusion that missing large-scale structure does not play a major
role in these observations. 

Therefore, the structure in polarized intensity must be due wholly to
depolarization mechanisms. For a pure Faraday screen the only kind of
depolarization that is possible is beam depolarization, because the
observed values of $RM$ imply that bandwidth depolarization is not
important, while depth depolarization requires that the rotating
medium emits as well. However, beam depolarization can only explain
structure in $P$ on beam-size scales. Therefore we are led to
consider the more realistic situation in which we observe a polarized
background that is modulated by a layer that both causes Faraday
rotation, and emits synchrotron radiation. Note that the argument
which limits the importance of offsets through the width of the
distribution of observed $RM$s applies equally to a pure Faraday
rotating screen and to a rotating and emitting screen.

The only way in which offsets could play a r\^ole is if there were a
layer in front of the rotating and emitting screen which emits
polarized radiation that is constant over the primary beam of our
observations.  A foreground-offset decreases the degree of
polarization with a constant factor and can contribute a constant $RM$
component which cannot be derived from the data.  However, a uniformly
polarized foreground cannot influence the width of observed $RM$
distribution or induce small-scale depolarization.  Judging from
earlier single-dish data of $RM$ in the regions of the Auriga and the
Horologium region (Bingham and Shakeshaft 1967, Spoelstra 1984), we
conclude that a possible undetected $RM$ component on scales $\ga 1\dg$,
if present at all, must be very small.

\section{Depolarization canals}
\label{s3:can}

A conspicuous feature in the observed polarized intensity is the
presence of one-dimensional filament-like structures of low polarized
intensity $P$.\@ These so-called depolarization canals have been
observed in many diffuse polarization observations (Wieringa et al.\
1993, Duncan et al.\ 1999, Gray et al.\ 1999, Uyan\i ker et al.\
1998, Gaensler et al.\ 2001). Two characteristics of the canals in
our observations (and in others as far as we could judge from figures)
are (1) the canals are one beam wide, and (2) the polarization angle
changes across the canal by 90\dg (Haverkorn et al.\ 2000). This
characteristic behavior can be explained by two mechanisms:
\begin{enumerate}
  \item The canals can denote a boundary between two regions which
        each have approximately constant polarization angle, but
        between which there is a difference in polarization angle of
        $\Delta \phi = (n + 1/2) 180\dg$ ($n = 0, 1, 2,\ldots$) will
        cause almost total depolarization through vector addition, if
        the polarized intensities on either side of the boundary are
        essentially identical (Haverkorn et al.\ 2000). 
        The resulting canal is, by definition, one beam wide. 
  \item A medium containing a uniform magnetic field, thermal
        electrons and cosmic-ray electrons depolarizes polarized
        radiation by means of differential Faraday rotation. In this
        case, the observed polarized intensity is
        \begin{equation} P = P_0 \;
           \left|\frac{\sin(2\,RM\,\lambda^2)}{2\,RM\,\lambda^2}\right|
   	   \label{e3:burn}
	\end{equation}
	(Burn 1966, Sokoloff et al.\ 1998), where $P_0$ is the
	polarized intensity observed at $\lambda \rightarrow 0$. A
	linear gradient in $RM$ would produce long narrow
	depolarization canals at a certain value $RM_c$ where $2
	\,RM_c \,\lambda^2 = n \pi$. Across every null in the
	sinc-function, the polarization angle changes by 90\dg.
\end{enumerate}
We will attempt to estimate the importance of each of the two
mechanisms in our observations.

\subsection{Frequency dependence of canals}

If the canals are due to beam depolarization, there are two extreme
possibilities for the origin of the angle change $\Delta \phi =
(n+1/2)\;180\dg$: it can be due to either an RM change across a canal of
$(n+1/2)\;180\dg/\lambda^2$, or to an intrinsic angle difference
of $\Delta\phi_0=(n+1/2)\;180$\dg. These two extremes cannot be
distinguished from observations at a single frequency. However, one
would expect $P$ in the canals to vary with frequency if the RM
changes across a canal, while the depth of a canal should be constant
for a $\Delta\phi_0$. If, on the other hand, the canals are caused by
differential Faraday rotation, $P$ in a canal should change with
frequency according to Eq.~(\ref{e3:burn}).

\begin{figure*}
  \centering
  \begin{center}
    \psfig{figure=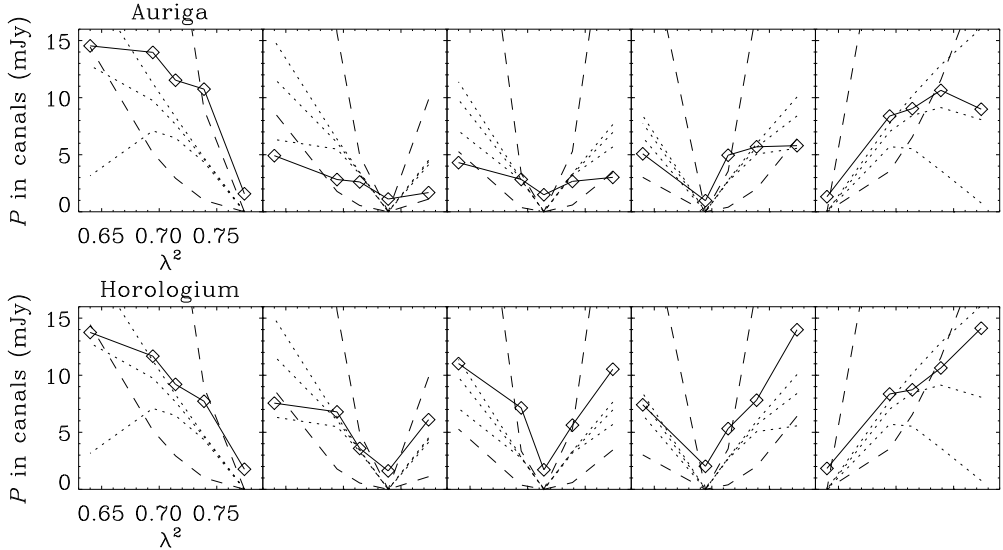,width=.75\textwidth}
    \caption{Frequency dependence of the depth of the canals, in the
      Auriga  region (top) and Horologium region (bottom). $P$ is the
      average over all canal-pixels, and in each plot the canals were defined
      in an other frequency band, at 341, 349, 355, 360, and 375 MHz
      from left to right. The prediction for $P(\lambda^2)$ if all
      canals were caused by beam depolarization due to a change in
      $RM$ is shown in dashed lines, from bottom to top for $\Delta RM
      \approx$ 2.1, 6.3 \radm\ (i.e.\ $\Delta\phi = 90\dg, 270\dg$). 
      The prediction for $P(\lambda^2)$ if the canals were caused
      by differential Faraday rotation are shown dotted for $2 RM_c
      \lambda^2 = n\pi$ for $n = 1, 3$ and~5.}
    \label{f3:can_l2}
  \end{center}
  \vspace*{0.5cm}
\end{figure*}

We have tested the frequency dependence of the depth of the canals as
follows. Canals are defined as sets of ``canal-pixels''. A pixel is
defined as a ``canal-pixel'' if the polarized intensity is low ($P <$
2 times rms noise) and $P$ on diametrically opposed sides of that
pixel, one beam away, is high ($P > 5$ times rms noise). The high-$P$
pixels surrounding the canal-pixel can be oriented horizontally,
vertically or diagonally. No further assumptions regarding the length
of canals are made; therefore a single pixel with low $P$ that is no
part of a canal but is surrounded by high $P$ pixels, is also defined
to be a canal-pixel.  

Sets of canal-pixels are evaluated for each frequency separately, so
that five sets of canal-pixels result. The average $P$ in each set of
canal-pixels is computed at all frequencies. In Fig.~\ref{f3:can_l2},
we plot the average values of $P$ against $\lambda^2$ for the five
sets of canal-pixels (where canals are defined at one of the
frequencies)  in the Auriga and Horologium regions. In each panel in
Fig.~\ref{f3:can_l2}, i.e.\ for each frequency, and both in Auriga and 
Horologium, the wavelength in which the canals are selected has the
lowest average $P$, with $P$ increasing with $|\Delta \lambda^2|$,
i.e.\ the canals decrease in depth at the other wavelengths. This
rules out the possibility that the canals are caused by a change in
intrinsic angle, confirming the conclusion from the non-detection of
$I$ that the background polarized intensity is smooth.

The dashed lines show the predictions of $P(\lambda^2)$ for canals
that are caused by beam depolarization, and are due to a change in
$RM$ (with $\Delta RM$~=~2.1 and 6.3~\radm, respectively). The dotted
lines in Fig.~\ref{f3:can_l2} denote the prediction of the
polarization angle if the canal is caused by differential Faraday
rotation, from Eq.~\ref{e3:burn}). Both predictions have arbitrary
polarized intensities. Therefore, the scaling of the models contains
no physical information, and is adjusted to fit the data. 

The accuracy of the model predictions can be judged by the shape of
the predicted lines, which is typical of the responsible depolarization
process. Furthermore, the same scaling should be used for each
frequency.  Judging solely from the shape of the lines, the
prediction of differential Faraday rotation seems to make a fit
somewhat better than that of beam depolarization, but not by
much. This is not totally unexpected, because it is probable that a
combination of both processes is at work in the majority of pixels.

\subsection{$RM$ and $\Delta RM$ values in canals}

\begin{figure*}
  \begin{center}
    \psfig{figure=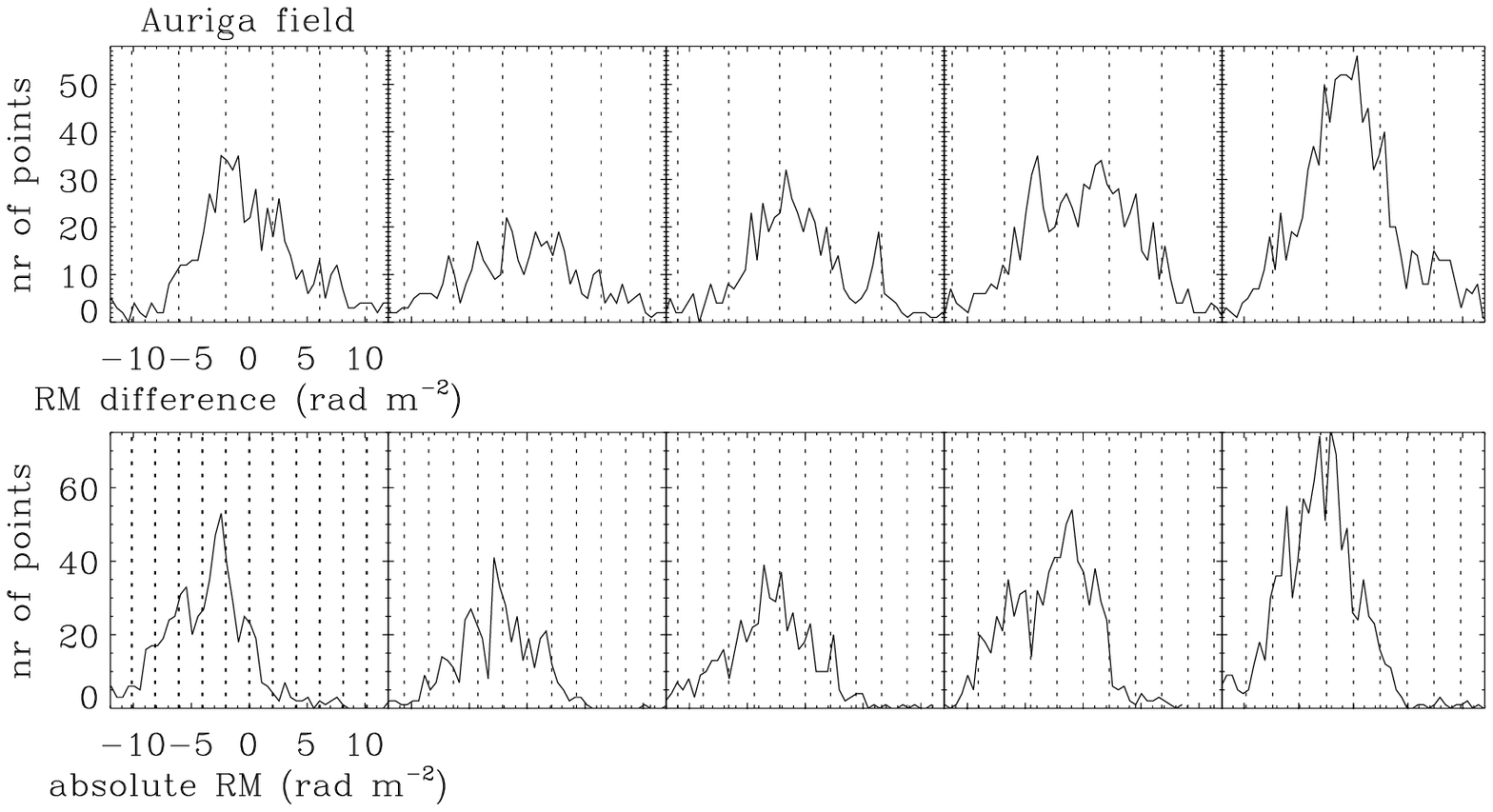,width=.75\textwidth}
    \psfig{figure=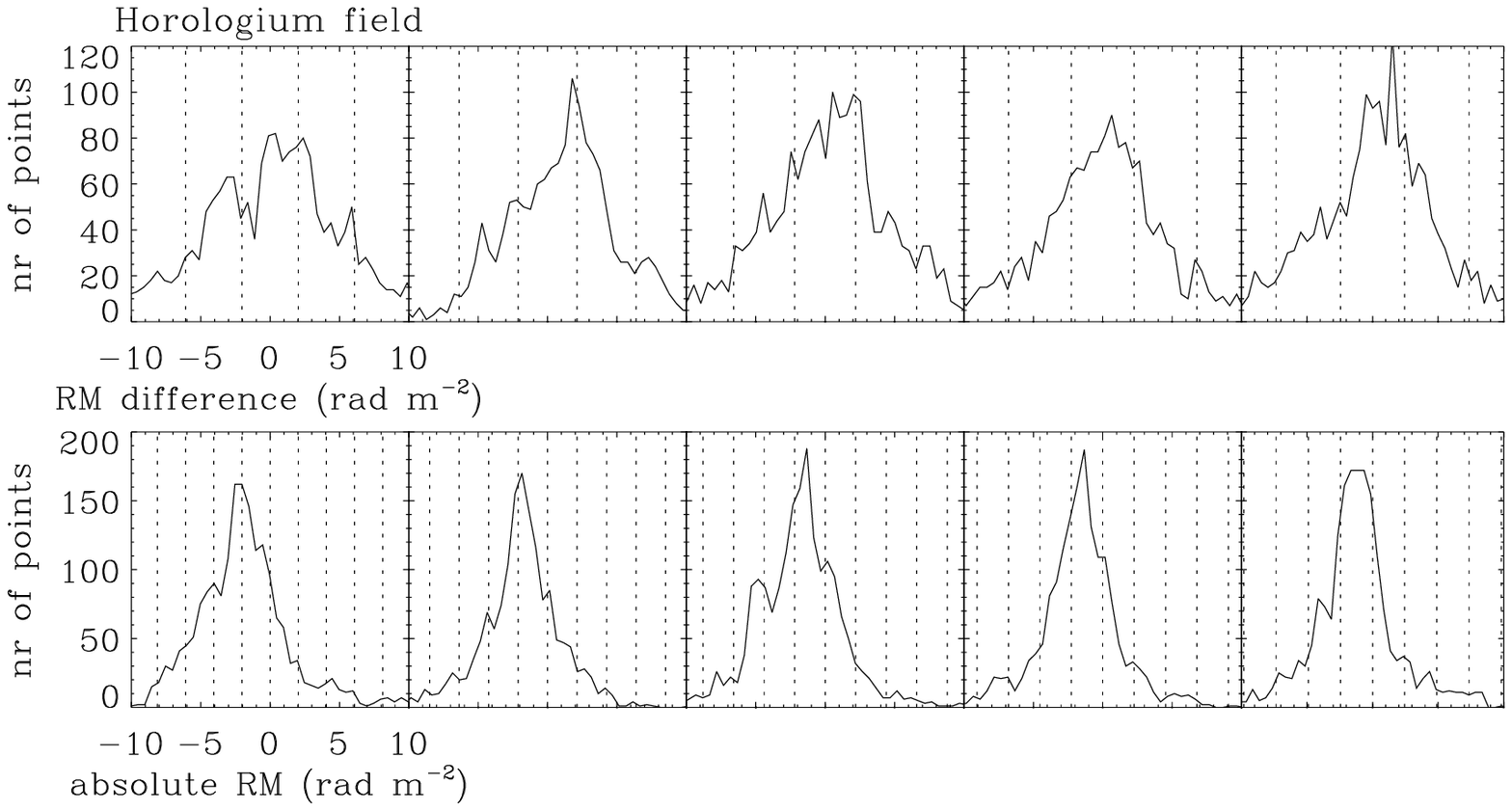,width=.75\textwidth}
    \caption{Distribution of $\Delta RM_c$ across canals and the
      absolute $RM_c$ at the canal position (as estimated from its
      neighbors). Canals are defined at frequencies 341, 349, 355, 360,
      and 375 MHz from left to right. Top panels show the Auriga
      region and bottom panels the Horologium region. Only $RM_c$
      values where $\chi^2_{red} < 2$ and $P > 20$~mJy/beam are
      used. In the $\Delta RM_c$ plots, dotted vertical lines are
      $\Delta RM_c$ values where $\Delta\phi$ across a canal would be
      $\pm 90$\dg, $\pm 270$\dg\ or $\pm 450$\dg. In the $RM_c$ plots,
      the dotted vertical lines are values where $2 RM_c \lambda^2 =
      n\pi$.}
    \label{f3:can_rm}
  \end{center}
\end{figure*}

Canals due to beam depolarization are caused by a specific change in
$RM$ across a canal $\Delta RM_c = (n+1/2) \pi/\lambda^2$. On the
other hand, canals caused by differential Faraday rotation are
determined by a specific absolute $RM_c = n \pi/ (2\lambda^2)$. With
the sets of canal-pixels defined in the previous subsection, we define
$\Delta RM = RM_1 - RM_2$, where 1 and 2 are high-$P$ pixels on
opposite sides of the canal. Then the RM at the canal-pixel is
estimated as $RM = (RM_1 + RM_2)/2$. The observed distributions of
$\Delta RM$ and $RM$ are shown in Fig.~\ref{f3:can_rm}. Both $\Delta
RM$ and $RM$ distributions show peaks at the values that
will produce canals, and the observations do not show perfect
agreement with either of them. Note that canals with angle changes
$\Delta\phi \la 90$\dg\ have accompanying $\Delta RM$s or $RM$s
 slightly different from canals with $\Delta\phi = 90$\dg\  and
therefore broaden the peaks. Noise in $RM$ has the same effect. 

\subsection{Position shift of the canals with frequency}

Differential Faraday rotation causes total depolarization at all
positions where $RM = RM_c$. This  means that depolarization 
not necessarily creates narrow one-dimensional canals, but
could also produce patches of $RM = RM_c$ that are larger than one
beam. The easiest way to explain one-dimensional canals in this 
picture is to assume a gradient in $RM$, so that the $RM$ stays 
constant over a certain length perpendicular to the gradient, and a
one-dimensional canal is formed.  But at 375~MHz, the canal will form
at positions where $RM_c$~=~2.46~\radm\, while  at 341~MHz it
forms where $RM_c$~=~2.02~\radm. If we assume typical gradients of
1~\radm\ per degree, similar to the large-scale gradient observed in
the Auriga region, this indicates that the canal should move with
position over $\sim$~5 beams from 341~MHz to 375~MHz. Instead, canals
move at maximum 3 pixels from 341 to 375~MHz, which is about
0.5~beam. {\em All} gradients in $RM$ would have to be larger than
 $\sim$~1~\radm\ per beam to position the canals in the 5
frequencies within half a beam from each other. Such a gradient is
certainly possible locally, although this high gradient would have to
extend over a large part of the field in order to explain the long and
straight canals.  Furthermore, if such large gradients are present in
the medium, we would expect lower gradients as well. These lower
gradients would give canals that shift position with frequency
significantly, which we do not observe.

\subsection{Shape of the canals}

\begin{figure}
  \begin{center}
    \psfig{figure=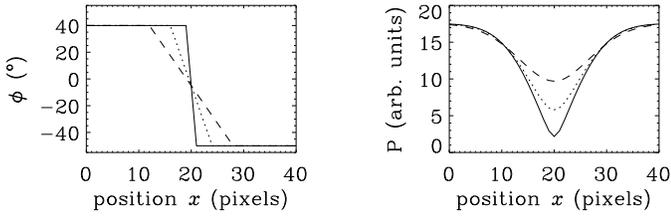,width=.5\textwidth}
    \caption{Predicted canal shape for a toy model with angle changes
      of 90\dg\ across a canal with different gradients (left). 
      Convolution of $Q$ and $U$ gives the $P$ distribution as shown
      right for the 3 gradients. The width of the beam is 12 pixels.}
    \label{f3:can_the}
  \end{center}
\end{figure}
\begin{figure}
  \begin{center}
    \psfig{figure=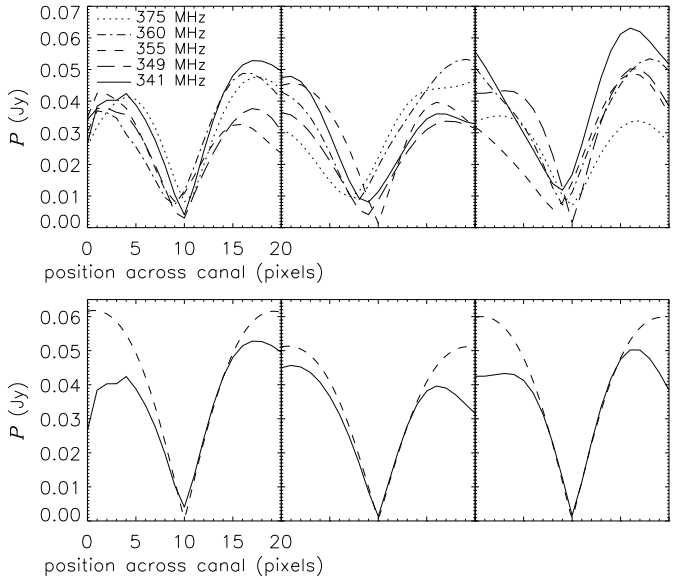,width=.5\textwidth}
    \caption{Measures shapes of the deepest canals. Upper plots:
      examples of observed one-dimensional $P$ distributions in the
      Horologium region for five frequencies, where the deepest canal
      is observed at 341, 355 and 349 MHz respectively. Lower plots:
      the same $P$ distribution for the deepest canal as above (solid)
      and the best fit according to the model of
      Fig.~\ref{f3:can_the}. The $P$ profile is so steep that the
      change in angle that causes the canal must be on scales of an
      arcminute or smaller.}
    \label{f3:can_shape}
  \end{center}
\end{figure}

The shape of the decline in $P$ across a canal, or the ``steepness''
of a canal, gives a lower limit to the abruptness of the change in
polarization angle across a canal, as can be seen in
Fig.~\ref{f3:can_the}. Here a one-dimensional example is given of a
change in polarization angle $\Delta\phi=90\dg$ (left) and the
corresponding change in $P$ after convolution with the telescope
beam (right). The narrowest $P$~profile is achieved when the change in
angle is on a length scale smaller than about one fifth of the beam. 

In Fig.~\ref{f3:can_shape}, the steepest canals found in the data are
shown. In this figure the top plots give a one-dimensional cross-cut
of $P$ across three canals against position, for all frequencies. The
frequencies in which the canals were defined were 341 MHz, 355 MHz and
349 MHz respectively, and the canals were selected for their
steepness. The bottom plots give only the $P$ distribution across the
canal at the frequency at which it was defined (solid line).
Superimposed in dashed lines is the $P$ distribution of the model of
Fig.~\ref{f3:can_the} for the steepest angle change convolved with the
synthesized beam. Less steep angle changes give less steep $P$
profiles and worse fits to the data. An interpretation of these
(specifically selected) steep canals in terms of differential Faraday
dispersion is difficult, because the canals would have to be much more
closely spaced than observed. Beam depolarization predicts a
change in depth of the canals across the frequency bands of about
20\%, in agreement with the observations.

\subsection{Canals due to beam depolarization}

We conclude that the dominant process creating one-beam wide canals of
almost complete depolarization is most likely beam depolarization.  In
this case, abrupt $RM$ changes have to be present in the medium. It
may seem fortuitous that only RM gradients of the right magnitude to
make canals would exist. However, this is not the case: RM gradients
of any magnitude are likely to occur in the medium, but only the RM
gradients that cause $\Delta\phi\approx 90\dg$ yield a visible
signature in $P$. Because $RM$ is an integral along the line of sight,
it is difficult to see what physical process would be responsible for 
this. However, numerical models of a magneto-ionized ISM show that
RM gradients steep enough to produce canals at 350~MHz are common
(Haverkorn \& Heitsch 2004). The relatively low RM gradient needed to
make a canal at 350~MHz, as compared to 1.4~GHz observations, could
explain why canals are abundant in these WSRT observations, but are
much less common at 1.4~GHz (Uyan\i ker et al.\ 1998). Nevertheless,
Figs.~\ref{f3:can_l2} and~\ref{f3:can_rm} show that beam
depolarization certainly is not the whole explanation.

If differential Faraday rotation were the main cause of the canals, it
would be hard to understand why all canals are exactly one beam wide,
and why we do not observe any significant change in the position of
the canals with frequency. Furthermore, the existence of canals in
which $P$ goes down to almost zero would then indicate a very uniform
medium in both magnetic field and electron density. Sokoloff et al.\
(1998) showed that an exponential asymmetric slab causes non-zero
minima for the canals, which even disappear completely in a turbulent
medium. Small-scale structure in observed $RM$ indicates that
small-scale structure in magnetic field and/or electron density is
abundant, so that a uniform medium needed for deep canals in the
differential Faraday rotation interpretation is unlikely. However,
Shukurov and Berkhuijsen (2003) argue that the canals they observed at
1.4~GHz in M31 are best explained as due to depth depolarization.

\section{Conclusions}
\label{s3:conc}

Small-scale structure in the linearly polarized component of the
diffuse Galactic synchrotron emission is seen in almost every
direction. Mostly, this structure is not correlated with total
emission, and therefore cannot be due to small-scale structure
in emission. Instead, the polarization angle $\phi$ is Faraday-rotated
in the magneto-ionic medium through which the linearly polarized
radiation propagates. However, the structure in polarized intensity
$P$ cannot be produced by Faraday rotation alone (which only rotates
$\phi$), but there are several other processes responsible for
this. First, instrument-related effects produce structure in $P$, such
as large-scale components in the radiation that are undetectable with
an interferometer, depolarization due to variation in angle within the
telescope beam, or over the frequency band width. Furthermore,
physical depolarization processes in the ISM can cause depolarization
if Faraday rotation and synchrotron emission occur in the same medium.

In this paper, we have discussed these processes and gauged their
relative importance in two sets of observations made with the
Westerbork Synthesis Radio Telescope (WSRT). 

Undetectable large-scale components in Stokes $Q$ and/or $U$
measurements can create structure in $P$, and prohibit the correct
determination of rotation measure. However, we showed that in our
fields, the observed range in rotation measure is so large that
offsets cannot play a significant r\^ole.

Narrow one-beam-wide canals of depolarization can be caused
by beam depolarization or differential Faraday rotation. Our
observations suggest that beam depolarization is the dominant
mechanism responsible for the canals at 350~MHz, although depth
depolarization is likely to contribute.

\section*{Acknowledgements}
We wish to thank R. Beck, E. Berkhuijsen and J. Tinbergen 
for helpful discussions.  The Westerbork Synthesis Radio
Telescope is operated by the Netherlands Foundation for Research in
Astronomy (ASTRON) with financial support from the Netherlands
Organization for Scientific Research (NWO). MH acknowledges support
from NWO grant 614-21-006.

\appendix 

\section{Offsets for a non-emitting Faraday screen}
\label{a3:screen}

First we consider the situation of a small-scale Faraday screen, i.e.\
of a constant polarized background emission that undergoes Faraday
rotation while propagating through a magneto-ionized medium. In this
case, small-scale structure in polarization angle is created by the
Faraday rotation, while the polarized intensity remains unaltered. We
assume a uniform polarization background ${\bf P_0} =
P_0\;\mbox{exp}(-2i\phi_0)$, where $\phi_0$ is the intrinsic
polarization angle. Assuming that the offsets can be approximated by a
constant over the whole field of observation, the expected offsets can
be derived depending on the $RM$ distribution in the screen. We
consider the case in which the Faraday screen consists of cells with
random $RM_r$ drawn from a Gaussian $RM$ distribution of width \srm\
and thus Faraday-rotates the background polarization angle on small
scales.  The offsets are then the normalized mean of the polarized
emission ${\bf P_0}=P_0\;\mbox{exp}(-2i\phi_0)$ weighted with the
Gaussian $RM$ distribution: 
\begin{eqnarray}
  {\bf P}_{\mbox{\em offsets}} &=& \frac{\int_{-\infty}^{\infty} P_0 \;
              \mbox{e}^{2i(\phi_0 + RM_r\lambda^2)} \;n(RM_r) \; dRM_r}
              {\int_{-\infty}^{\infty} n(RM_r)
              \; dRM_r} \\
	      && \hspace*{-1cm}\mbox{where} \;\;\; 
	      n(RM_r) = \mbox{e}^{-RM_r^2/2\srm^2} \nonumber 
\end{eqnarray}
This expression is independent of the angular length scale of the
structure in $RM$, as long as the length scale is small enough with
respect to the path length to have a Gaussian distribution of $RM$s.
This yields the offsets $Q_0$ and $U_0$
\begin{eqnarray}
  Q_0  &=& A \left[ \cos(2\phi_0) - \frac{1}{\sqrt{\pi}}
             \sin(2\phi_0)\;F\right] \label{ea:q0}\\
  U_0  &=& A \left[ \sin(2\phi_0) - \frac{1}{\sqrt{\pi}}
             \cos(2\phi_0)\;F \right] \label{ea:u0}\\
  \mbox{where}\;F&=& \int_0^{\sqrt{2}\srm\lambda^2} \mbox{e}^{t^2} dt 
              \;\;\;\mbox{and}\;\;\;
	     A= P_0 \mbox{e}^{-2\srm^2\lambda^4} \nonumber
\end{eqnarray}
The offsets depend highly non-linearly upon the width of the random $RM$
distribution~\srm. The exact values can be easily calculated
analytically for two extremes:
\begin{itemize}
  \item[a)] The width of the $\phi$-distribution $\sigma_{\phi} =
  \srm\lambda^2 \ga \pi$, or large $\srm$:
	   \[
              \mbox{e}^{-2\srm^2\lambda^4}
              \rightarrow 0 \;\;\;\; \Rightarrow \;\;\;\; Q_0 = U_0 = 0
              \]
	   The observed distribution of polarization angles is random,
           therefore $Q$ and $U$ are centered around zero and there is
           no undetected large-scale structure.
  \item[b)] \srm\ and accompanying $\sigma_{\phi}$ are so small
           that $\srm\lambda^2\ll 1$:
           \[
	      \mbox{e}^{-2\srm^2\lambda^4} \rightarrow 1
	      \;\;\;\; \Rightarrow \;\;\;\;
	      \left\{ \begin{array}{lll} 
		Q_0 & = & P_0\;\cos(2\phi_0) \\ 
		U_0 & = & P_0\;\sin(2\phi_0) \\
	      \end{array}  \right. 
	   \]
	   Here the subtracted component is equal to the uniform
           component of the polarization vector. The observed
           polarized intensity is much lower than the true polarized
           intensity because of these large offsets.
\end{itemize}
A constant background rotation measure $RM_u$ can be incorporated by
replacing $\phi_0$ in the equations by $\phi_u = \phi_0 + RM_u
\lambda^2$, but this does not change the results.

\end{document}